\documentclass[acmsmall,screen,nonacm]{acmart}
\usepackage{booktabs,tabularx,array}
\newcolumntype{Y}{>{\raggedright\arraybackslash}X}
\newcommand{\sut}{\textsc{sut}}
\title[LLM-Based Test Generation]{LLM-Based Test Generation: Information Sources, Generation Strategies, and Quality Evidence}
\author{Yunhao Liang}
\affiliation{\institution{Chengdu Institute of Computer Applications, Chinese Academy of Sciences}\country{China}}
\affiliation{\institution{University of Chinese Academy of Sciences}\country{China}}
\author{Chengguang Gan}
\affiliation{\institution{Independent Researcher}\country{Japan}}
\author{Ruixuan Ying}
\affiliation{\institution{Institute of Multidisciplinary Research for Advanced Materials (IMRAM), Tohoku University}\country{Japan}}
\author{Hanjun Wei}
\affiliation{\institution{University of Chinese Academy of Sciences}\country{China}}
\author{Zhe Cui}
\affiliation{\institution{Chengdu Institute of Computer Applications, Chinese Academy of Sciences}\country{China}}
\affiliation{\institution{University of Chinese Academy of Sciences}\country{China}}
\author{Shiwen Ni}
\affiliation{\institution{Artificial Intelligence Research Institute, Shenzhen University of Advanced Technology}\country{China}}

\hypersetup{pdfauthor={Yunhao Liang, Chengguang Gan, Ruixuan Ying, Hanjun Wei, Zhe Cui, Shiwen Ni}}
\begin{document}
\begin{abstract}
Large language models are increasingly used to generate test scenarios, executable test suites, assertions, interaction sequences, and fuzzing infrastructure. These artifacts serve different purposes and rely on different sources of information about correct behavior. A test that increases implementation coverage, a test that agrees with a reference program, and a test that detects a requirements violation therefore provide distinct kinds of evidence. We present a structured narrative survey that connects the generation process to the evidence used to justify test quality. Drawing on 95 curated source records through 29 September 2026, we organize the field along four dimensions: testing objectives and artifacts, information available during training and generation, generation and learning mechanisms, and evaluation evidence. We synthesize work on unit and requirements-based testing, test oracles, API and GUI testing, fuzzing, and learned test generators. The resulting framework explains how execution feedback can improve executability while also affecting the independence of an oracle, why specification provenance and training-time reference supervision matter for comparisons, and how downstream code-selection gains differ from test correctness. We use these distinctions to organize benchmarks and derive a research agenda for independent oracle assessment, budget-aware generation, repository-scale evaluation, and test maintenance.
\end{abstract}
\maketitle
\section{Introduction}
Test generation requires a connection between what software should do and observations that can expose when it does something else. Constructing that connection involves several kinds of work: interpreting requirements, selecting inputs, arranging an execution environment, choosing observations, and expressing an expected result. Large language models (LLMs) can contribute to each of these activities because they operate across natural language, program code, and semi-structured technical documentation. Research has consequently expanded from generating unit-test code to producing acceptance scenarios, postconditions, API call sequences, and fuzzing drivers~\cite{S01,S02,S05}.

This breadth creates a comparison problem. Two approaches can both generate executable tests while having different access to the system under test (\sut), different sources of expected behavior, and different intended uses. A regression generator may inspect an implementation and retain tests that pass on it. A requirements-driven generator may need to produce an oracle before an implementation exists. An issue-reproduction agent may receive a report of an already observed failure. The same passing test carries a different meaning in each setting. Comparing their coverage or execution success without describing these conditions obscures what has actually been learned about test quality.

TestGen-LLM filters regression improvements through execution and coverage~\cite{U07}; CoverUp uses coverage to guide generation~\cite{U05}; PyTester trains description-conditioned generation using reference execution~\cite{T01}; UTRL couples code and test generators through adversarial learning~\cite{T02}. These methods differ in when behavior becomes available and how it influences tests.

Existing surveys provide substantial foundations. Broad reviews map LLM applications across testing and repair~\cite{S01,S04}; unit-testing surveys organize techniques, model adaptation, and quality assurance~\cite{S02,S03}; requirements-based reviews cover transformations from natural-language and formal specifications~\cite{S05,S06}. A recent systematic review directly addresses LLM-based test-case generation~\cite{S07}. Oracle generation, hybrid techniques, and evaluation limitations already appear in these accounts. Our organizing question is how these aspects interact: \emph{what information produces a test, and what evidence justifies using that test to judge software?}

We develop this question through a structured narrative synthesis of a curated collection of 95 source records. The collection includes seven surveys and roadmaps, together with primary methods, benchmarks, empirical studies, and selected neural predecessors. Its breadth supports a cross-setting analysis; its preliminary screening status is described in Section~\ref{sec:scope}. We make three contributions. First, we define a framework that separates testing objectives, generated artifacts, information access, and quality evidence. Second, we use the framework to connect prompting, program analysis, feedback, agents, and model training across testing settings. Third, we organize evaluation designs according to the claims they can support and derive concrete research questions from the remaining mismatches.

The central argument is that test generation should be understood as both an artifact-construction problem and an evidence-construction problem. Better syntax and broader execution improve the first. Confidence in expected behavior requires the second. This distinction helps explain when tools from different research traditions can be combined and which claims require additional validation.

\section{Foundations and Analytical Framework}
\label{sec:framework}
\begin{figure}[t]
\centering
\includegraphics[width=\linewidth]{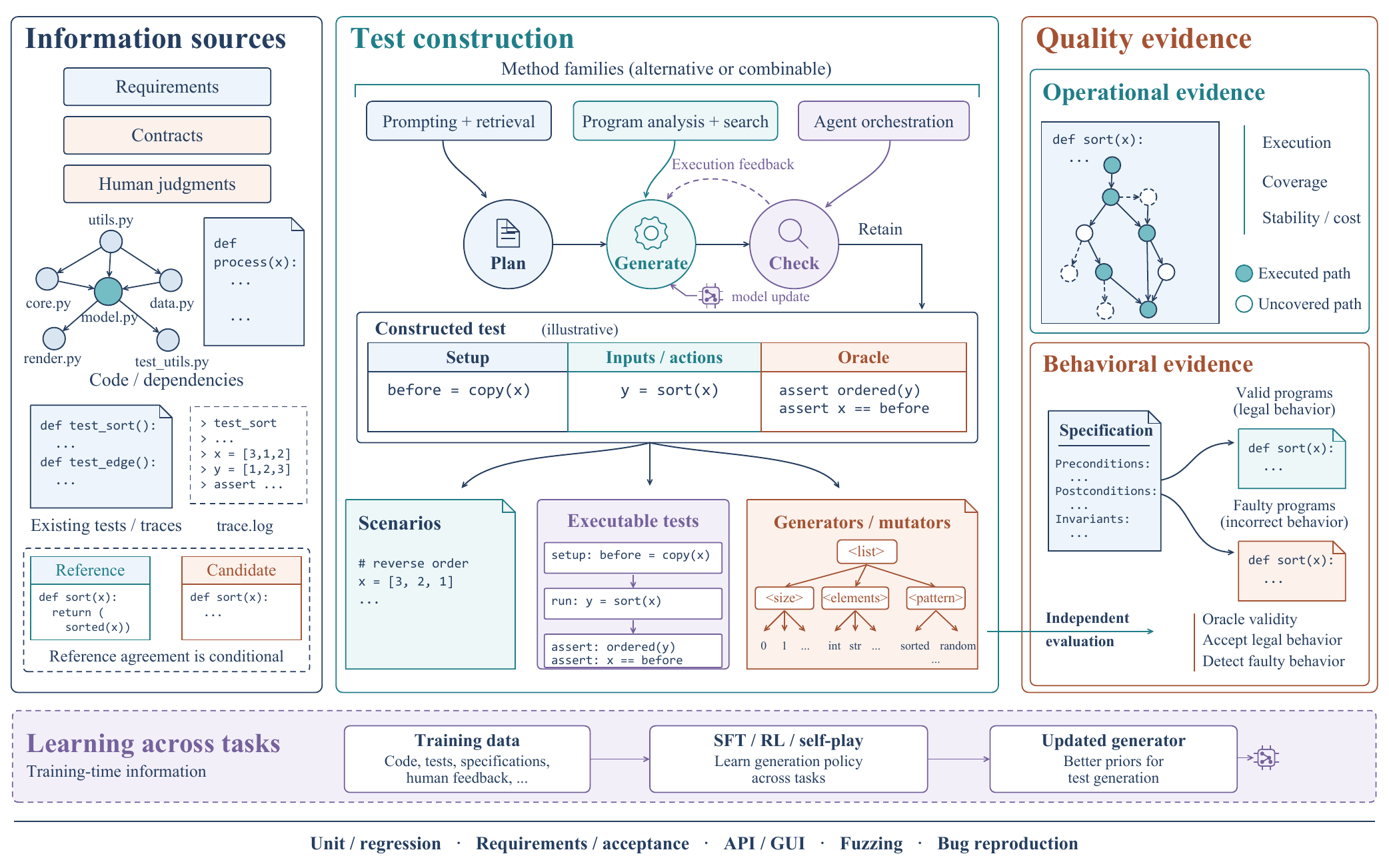}
\caption{A synthesis of LLM-based test generation. Information sources support planning, generation, and checking; alternative or combined mechanisms construct scenarios, executable tests, and generators. The sorting sketch illustrates setup, action, and a partial oracle. Dashed feedback revises the current artifact; matching violet model icons connect the cross-task learning output to generation. Operational and behavioral evidence address different quality properties. Independent evaluation supplies held-out evidence without feeding its results back into generation.}
\Description{Three separate requirement, contract, and human-judgment cards appear above code dependencies, existing tests and traces, and reference and candidate programs. Planning, generation, and checking construct an illustrative sorting test with setup, action, and oracle. Three output branches show scenarios, executable tests, and generators. A separate training lane supplies an updated model, represented by matching model-chip icons. Operational path evidence and independent behavioral evaluation occupy the right-hand panel.}
\label{fig:overview}
\end{figure}
\subsection{What Is Generated?}
An executable test combines a setup, an input or action sequence, and an oracle that judges the resulting behavior. Its setup can include fixtures, mocks, dependencies, credentials, database state, and cleanup. Its input may be a function argument, an HTTP request, a GUI action, or an entire generated program. Its oracle may check an exact output, an exception, a relation across executions, or a safety property. A test suite additionally determines which tests coexist, how state is isolated, and when execution ends.

Many generation methods produce an intermediate artifact. A natural-language scenario still requires operationalization. A grammar defines a family of admissible inputs but does not necessarily determine their expected outputs. A fuzzing driver connects generated data to a target API while relying on a separate failure detector. We therefore distinguish \emph{test design}, \emph{test input}, \emph{test oracle}, \emph{executable test}, and \emph{test-generation infrastructure}. This distinction preserves the contributions of input-only and infrastructure methods without attributing complete test synthesis to them.

\subsection{Testing Objectives Determine Oracle Roles}
Functional conformance asks whether behavior satisfies an external requirement. Regression testing asks whether a change violates behavior that should be preserved. Bug reproduction asks whether a reported failure can be exercised and distinguished from its repair. Fuzzing often searches for crashes, sanitizer violations, or disagreement among implementations. Tests used in code generation may instead rank candidate solutions or provide debugging feedback. These objectives can share execution machinery while requiring different oracle evidence.

Consider a requirement that a function return a sorted permutation of its input without modifying the input. A test may assert the sorted result yet omit the mutation constraint. It can execute successfully and reach every line while leaving part of the requirement unchecked. A test copied from an implementation that sorts in place can even preserve the very violation that a conformance test should reveal. Conversely, a test that expects one particular ordering among equal-key objects may reject a legal implementation if stability was never required. This illustrative example motivates evaluating both detection of invalid behavior and acceptance of permitted behavior.

\subsection{Information Access and Correctness Evidence}
We distinguish three stages of information access: training, generation, and final evaluation. At each stage, relevant sources include requirements, focal code, repository context, existing tests, reference implementations, candidate implementations, execution traces, and human decisions. Access is a property of an experimental variant rather than an entire paper. For instance, inference can use a description and signature even when training rewards require a reference program, as in PyTester~\cite{T01}. Describing that configuration simply as ``code-free'' would erase a material dependency.

An information source can play more than one role. Source code helps resolve imports and construct objects; the same code can also supply expected outputs. A static dependency graph can identify callable APIs without validating their semantics. Execution of a trusted reference provides a concrete output under the reference assumption. Execution of the candidate under test provides an observation about that candidate. These roles should be recorded separately because context relevance and oracle independence answer different questions.

Specification provenance adds a fourth consideration alongside the three access stages. A docstring generated from buggy code remains implementation-derived, even if the next generator receives only that docstring~\cite{E06}. Conversely, a requirements-oracle study gives the generator requirements and minimal method structure, while researchers construct those requirements from buggy/fixed differences~\cite{O12}. Generation permissions and dataset-construction permissions are therefore separate fields. A PR's stated intent provides another behavioral source, interpreted alongside the change and before/after executions in Testora~\cite{R05}. These sources can all be useful, but their independence from the behavior under examination differs.

Figure~\ref{fig:overview} connects the four dimensions of our analytical framework: objectives and artifacts, information access, generation mechanisms, and quality evidence. The dimensions are intentionally orthogonal. ``Agent-based'' describes how decisions are coordinated; ``requirements-based'' describes an information source; ``unit testing'' describes a testing level. Treating these labels as competing branches of one taxonomy would place overlapping studies in arbitrary categories. We instead use testing settings to organize the landscape and mechanisms to explain recurring design choices.

\section{Review Scope and Synthesis Method}
\label{sec:scope}
\subsection{Research Questions}
Six questions guide the synthesis and connect its sections:
\begin{description}
\item[RQ1: Objectives and artifacts.] Which testing objectives, systems, and levels are addressed, and what artifacts are generated? Section~\ref{sec:landscape} maps the settings and outputs.
\item[RQ2: Information and oracles.] What information is available during training and generation, and where does evidence of correct behavior originate? Sections~\ref{sec:framework} and~\ref{sec:learning} distinguish access stages and sources.
\item[RQ3: Generation strategies.] How are LLMs combined with retrieval, analysis, execution, search, and agents? Section~\ref{sec:generation} compares these mechanisms.
\item[RQ4: Learning objectives.] What do training data, rewards, and adversarial curricula optimize? Section~\ref{sec:learning} examines learned generators.
\item[RQ5: Evaluation evidence.] Which quality claims do benchmarks support, and under what conditions are results comparable? Section~\ref{sec:evaluation} analyzes protocols and measurements.
\item[RQ6: Practical use and open problems.] What is supported by development-workflow evidence, and which capabilities still need validation? Section~\ref{sec:agenda} connects adoption, maintenance, and research directions.
\end{description}

\subsection{Collection and Selection}
This survey is a structured narrative synthesis of a preliminary literature collection assembled through 29 September 2026. Discovery used public web search, primary publication pages, arXiv records, author-hosted resources, and references identified through existing surveys. Query families combined large language models with unit-test generation, requirements-based testing, oracle generation, API and GUI testing, fuzzing, reinforcement learning, and evaluation benchmarks. Named-method searches expanded these families to identify related versions and recent developments. A supplemental pass added 35 records through topic-specific queries and targeted tracing of review bibliographies; its search notes and source-specific inspection records accompany the register. The source register contains 95 records, including seven surveys and roadmaps and 88 method, benchmark, empirical, and historical records; these counts describe the curated register rather than an exhaustive population of studies.

We include methods producing testing artifacts and studies evaluating their generation. Tests for debugging and code selection are included when they clarify test quality's downstream role. General code generation and testing LLMs themselves fall outside the central scope. Selected neural predecessors provide context~\cite{U11,O04}. Public primary sources were preferred. Retrieval excluded MDPI sources and records with unclear provenance, limiting coverage.

Related preprint and publication versions were treated as a study family. A changed title alone was not counted as a new method. The bibliography retains the inspected primary source and uses verified publication metadata where available. Several records remain preprint or author-resource citations even when their accompanying register reports an accepted venue; final proceedings normalization is incomplete. Publication status and evidence strength remain separate: a peer-reviewed abstract can establish a method's broad objective, while its detailed information access may still require inspection of the full text.

\subsection{Extraction and Interpretation}
The register records artifacts, information sources, mechanisms, analytical use, and inspection depth. Each entry was checked against a primary page or document. Most checks concern metadata and abstracts; selected passages establish information access, reward supervision, and metric definitions. We do not pool effect sizes or rank methods. Independent dual screening, inter-rater validation, and experimental reproduction have not been performed.

We distinguish method descriptions, interpretations of measured properties, and evaluation recommendations. Reference checking describes a method; reference consistency interprets its evidence; testing alternative legal implementations extends the protocol. Recommendations are not reported findings.

AI assistance was used to expand search terms, organize retrieved literature, propose the cross-study categories, draft the synthesis, and design its conceptual figures. The accompanying register preserves source links and inspection depth. The present coding has not received independent human validation. This status is relevant because category assignments can affect the survey's conclusions even when individual bibliographic records are correct.

\subsection{Relationship to Existing Reviews}
\begin{table}[t]
\caption{Existing review families and the emphasis of this synthesis. These are differences in scope and organization, not claims that individual topics were previously absent.}
\label{tab:surveys}
\small
\begin{tabularx}{\linewidth}{@{}p{.24\linewidth}YY@{}}
\toprule
Review family & Established coverage & Emphasis here \\
\midrule
Broad testing surveys and roadmap~\cite{S01,S04} & Testing tasks, LLM techniques, challenges, repair-related applications & Connect generated artifacts to their oracle evidence \\
Unit-testing surveys~\cite{S02,S03} & Generation methods, model adaptation, quality assurance, lifecycle concerns & Compare code-conditioned and specification-conditioned settings \\
Requirements-based reviews~\cite{S05,S06} & Requirements representations, transformations, industrial use & Separate scenario design, input construction, and executable oracles \\
LLM test-case SLR~\cite{S07} & Techniques, data, metrics, preprocessing and postprocessing & Track information access across training, inference, and assessment \\
\bottomrule
\end{tabularx}
\end{table}

The updated unit-test review by Chu et al. reports 178 publications eligible for full-text review through July 2026 in its September 2026 revision~\cite{S03}. Its count and this survey's 95-record register use different scopes and collection protocols. We therefore position our contribution through the connections in Table~\ref{tab:surveys}, rather than through a claim of more extensive retrieval.

Table~\ref{tab:surveys} places this account alongside existing reviews. Its intended value is the connection between dimensions: how a feedback source affects oracle independence, how the generated artifact changes the evaluation denominator, and how training supervision constrains claims of autonomy. These connections also provide a common vocabulary for integrating 2026 work on adversarial training, agents, and test maintenance into the earlier literature.

\section{Testing Settings and Generated Artifacts}
\label{sec:landscape}
\begin{figure}[t]
\centering
\includegraphics[width=\linewidth]{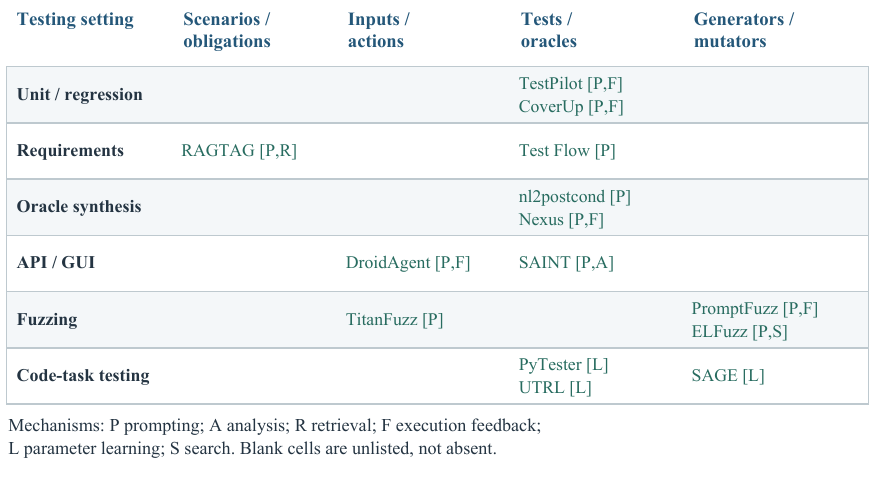}
\caption{A representative literature map, based on the methods discussed in this section~\cite{U01,U05,R01,R03,O01,O03,A02,A03,F01,F05,F06,T01,T02,T08}. A setting can produce several kinds of artifacts, and mechanisms can be combined. Blank cells indicate no example selected for display.}
\Description{A matrix locates representative methods across six testing settings and four artifact families. Mechanism labels distinguish prompting, analysis, retrieval, feedback, parameter learning, and search. Empty cells do not establish a research gap.}
\label{fig:literature-map}
\end{figure}

Figure~\ref{fig:literature-map} maps representative work to testing settings and generated artifacts. The mechanism labels identify selected operations discussed below; they are illustrative assignments rather than exhaustive coding or publication counts.

\subsection{Unit Tests and Repository Context}
Unit-test generation is the most visible setting in the reviewed collection. The task appears local because the target is a function or method, but executable tests depend on surrounding types, imports, initialization, and test conventions. AthenaTest established an earlier transformer-based approach that uses focal context and supervised code--test data~\cite{U11}. TestPilot combines function information and documentation with execution-based repair~\cite{U01}, while ChatUniTest organizes generation, validation, and repair into a framework~\cite{U02}. These approaches illustrate the transition from generating a plausible test body to constructing an executable artifact in its environment.

CITYWALK addresses project dependencies and C++ knowledge~\cite{U08}, exposing construction, linking, fixtures, and external dependencies absent from standalone-function tasks. Existing tests additionally supply conventions and an improvement baseline in TestGen-LLM~\cite{U07}. Integration value and the validity of inherited behavioral assumptions remain distinct.

ASTER makes static context and mocking part of test construction for Java and Python, while ChatTester combines generation with an iterative refinement process~\cite{U12,U18}. KTester distinguishes knowledge used in test design from knowledge used in implementing the test~\cite{U15}. Across these systems, framework familiarity, dependency resolution, and behavioral selection address different obstacles; an executable test need not be a well-chosen test.

\subsection{Requirements, Scenarios, and Acceptance Tests}
Requirements-driven generation begins with intended behavior expressed independently of an executable test. RAGTAG retrieves domain information to generate test scenarios from natural-language requirements~\cite{R01}. Multi-step generation of system-level test specifications uses intermediate artifacts to structure this transformation~\cite{R02}. Both settings expose an important distinction: producing a useful test design and producing a runnable script are separate accomplishments, with different evaluation methods and human responsibilities.

An industrial acceptance-testing study makes this separation explicit. AutoUAT transforms user stories into Gherkin scenarios, while Test Flow uses those scenarios and page HTML to construct Cypress scripts~\cite{R03}. Information access changes between stages. The scenario can be judged against a user story; the script also needs knowledge of concrete page structure. APITestGenie similarly combines requirements and API specifications to generate web API tests~\cite{R04}. These pipelines suggest that traceability should connect the requirement, the scenario, the executable actions, and the assertion, so that a failure can be attributed to the appropriate transformation.

Change intent and industrial tickets add intermediate settings between independent requirements and code-conditioned regression. Testora uses PR intent with code changes, existing tests, and before/after behavior to identify regressions~\cite{R05}. AToMIC combines JIRA tickets and recent code changes to construct Gherkin scenarios and mobile UI tests in an industrial case study~\cite{R06}. Intent traceability and executable-script construction should be assessed separately; the availability of natural-language text does not by itself establish implementation-independent generation.

\subsection{Oracles, Postconditions, and Metamorphic Relations}
Oracle generation isolates the part of a test that decides whether behavior is acceptable. TOGA generates exception and assertion oracles using a neural approach~\cite{O04}. TOGLL also learns assertion and exception oracles with code-conditioned prompts~\cite{O07}. TOGA's subsequent large-scale evaluation and replication work demonstrates the importance of examining oracle generation under changed evaluation conditions~\cite{O05}. An oracle should therefore be assessed by the behavior it accepts and rejects, in addition to whether it resembles a developer-written assertion.

Postconditions and metamorphic relations broaden the oracle representation. The nl2postcond work translates natural-language intent into executable method postconditions~\cite{O01}. Work on executable metamorphic relations derives relations from requirements and API specifications~\cite{O02}. A relation can test consistency across executions when predicting one exact output is difficult. Its applicability conditions are part of its meaning: a plausible transformation can be invalid for inputs that violate an unstated precondition. Thus, generating the relation and establishing its domain of validity remain distinct tasks.

Documentation can supply more explicit behavioral constraints. Doc2OracLL examines Javadoc components in oracle generation, whereas Tratto combines source and documentation with constraints on the tokens available for constructing axiomatic Boolean oracles~\cite{O08,O09}. Documentation selection concerns evidence relevance; grammatical and API constraints concern admissible expressions. Neither operation alone proves that a generated constraint captures the intended contract.

Nexus uses execution-grounded multi-agent processes for oracle synthesis~\cite{O03}, and CANDOR incorporates multi-agent consensus into end-to-end JUnit generation~\cite{O06}. These approaches combine test construction and oracle review. Their interaction patterns can expose contradictions, but the source of behavioral evidence still determines what agreement establishes. Agreement among agents or with candidate executions must be interpreted relative to the requirements being tested.

\subsection{API and GUI Interaction Sequences}
For APIs and GUIs, test generation must maintain state across actions. An endpoint can require a resource identifier returned by an earlier request; a GUI action can depend on a screen reached through navigation. AutoRestTest combines LLM support with multi-agent reinforcement learning for REST API testing~\cite{A01}. SAINT uses program analysis and LLM-based agents for service-level integration tests~\cite{A02}. LlamaRestTest adapts smaller models to API dependencies and input values using specifications and server responses, while ARMeta constructs executable metamorphic API tests from specifications~\cite{A05,A06}. The relevant artifact is therefore a coordinated interaction sequence, including dependencies between operations, rather than a collection of independent input values.

DroidAgent and GPTDroid apply LLM reasoning to mobile GUI exploration and interaction~\cite{A03,A04}; ScenGen coordinates observation, decision, execution, supervision, and recording around mobile scenarios~\cite{A07}. Here, observations and actions are interleaved: the next action depends on the current interface. Reaching a screen demonstrates exploration; checking that the screen implements a requirement additionally needs an oracle. This distinction also affects replay. A sequence generated through adaptive interaction may need stable selectors, state initialization, and checkpoints before it becomes a repeatable regression test.

\subsection{Fuzzing Inputs and Infrastructure}
Fuzzing gives the LLM a different role. TitanFuzz and FuzzGPT generate programs for testing deep-learning libraries, with the latter emphasizing edge cases~\cite{F01,F02}. Fuzz4All uses LLMs in a broader fuzzing framework~\cite{F03}. These methods exploit the model's ability to produce structured inputs while relying on execution and external failure detectors to identify anomalies.

The generated object can also be an input generator. ChatAFL extracts protocol-related information to support grammar construction and message-sequence exploration~\cite{F04}. PromptFuzz generates fuzzing drivers~\cite{F05}; ELFuzz searches over synthesized fuzzers~\cite{F06}; Bulbasaur generates customized mutators using branch-oriented information~\cite{F07}. G2FUZZ synthesizes reusable generators for non-textual inputs such as binary files, extending this distinction beyond generated textual programs~\cite{F08}. Scheduzz additionally couples API-combination constraints with driver generation and campaign scheduling~\cite{F10}. This progression changes the cost model because one generated artifact can produce many executions. A survey that records only the number of LLM-generated test cases would miss both the reuse and the contribution of the conventional fuzzing engine.

\subsection{Bug Reproduction and Tests for Code Generation}
LIBRO generates bug-reproducing tests from natural-language reports~\cite{G01}. SWT-Bench evaluates test generation in the context of real-world fixes~\cite{B03}. These settings provide a particular behavioral distinction: the test should expose the issue in a buggy version and accept a fixed version. Reproducing a known failure and discovering an unknown defect are related capabilities but use different information.

Generated tests also influence which code is selected or how it is repaired. CodeT uses generated tests to help choose among candidate programs~\cite{T10}, UTGen learns tests for automated debugging~\cite{T05}, and CodeRM scales test generation for code reward modeling~\cite{T06}. TiCoder introduces user feedback on tests to clarify intent during code generation~\cite{T11}. In all four cases, the downstream outcome depends on both the generator and the consumer of its tests. A useful survey must therefore retain test-level measurements alongside code-selection or repair outcomes.

\section{Generation Mechanisms}
\label{sec:generation}
\subsection{Context Construction and Task Decomposition}
Prompting determines which parts of a testing problem the model sees at once. Direct prompts usually combine a target, an instruction, and a desired output format. Examples can provide the test framework, naming conventions, and assertion style. Context construction becomes a substantive method contribution when it selects information needed to resolve dependencies or infer the intended behavior. TestPilot and ChatUniTest illustrate focal-code and environment-aware construction~\cite{U01,U02}; RAGTAG illustrates retrieval of domain material for requirements interpretation~\cite{R01}.

Structural context construction complements retrieval. HITS slices methods to organize generation around smaller code regions; TELPA incorporates program analysis and usage constraints to target difficult coverage; KTester supplies project and testing knowledge~\cite{U13,U14,U15}. Slicing reduces the local reasoning unit, analysis identifies relationships outside it, and domain knowledge supplies conventions or behavioral rules. Their contributions should be compared at the obstacle each addresses.

Retrieval has two separable functions. It can supply technical facts needed for execution, such as an API signature, or behavioral facts needed for an oracle, such as a business rule. These should be evaluated independently. An increase in compilation success can establish that retrieved API facts were useful without establishing that the retrieved business rule was current or applicable. A retrieval study on ML/DL APIs reports coverage improvements without improved correctness under its evaluation~\cite{U17}. This provides conditional evidence that technical reach and behavioral judgment can respond differently to added context. Doc2OracLL's documentation analysis and the context comparisons in a Java-oracle empirical study address related questions under different tasks~\cite{O08,E07}; their outcomes should not be pooled into a universal ranking of retrieval sources.

SymPrompt decomposes prompting around execution paths~\cite{U03}; requirements pipelines separate interpretation, scenario construction, and scripting~\cite{R02,R03}. Intermediate artifacts expose omitted scenarios, invalid locators, and wrong expectations as different errors. Their utility depends on final behavior as well as stage-specific success, because an early misunderstanding can propagate.

\subsection{Input, State, and Environment Construction}
Input construction must satisfy both the external input domain and the execution interface. Grammar-based methods such as SAGE and LogiCase make admissible input structures explicit~\cite{T08,T09}; repository-oriented approaches must also recover fixtures and dependencies~\cite{U08}. For API and GUI testing, validity extends to action order and state: a well-formed request can still be unusable before a required resource is created~\cite{A01,A02,A03}. These mechanisms solve different parts of reaching a meaningful observation.

Exception tests make the setup obligation concrete: exLong conditions specialized generation on throw traces, guards, and existing non-exception tests~\cite{U16}. ASTER's mocking addresses dependencies that would otherwise block execution~\cite{U12}. Reaching an exception and choosing whether that exception is expected remain separate decisions.

For the sorting example, preserving a copy of the original input makes mutation observable. Sorted expected values alone omit this setup obligation. Rejected inputs, failed setup, and assertion failures should therefore have separate denominators.

\subsection{Oracle Construction and Representation}
An oracle can predict an exact result, check a property, relate several executions, or obtain a human judgment. Postconditions and executable metamorphic relations express behavioral constraints~\cite{O01,O02}; TiCoder obtains user feedback to clarify intended behavior~\cite{T11}; Nexus and CANDOR use agent interaction in oracle construction~\cite{O03,O06}. The representation determines which behaviors the test can express, while the evidence source determines why those behaviors should be accepted.

For sorting, ordering, permutation preservation, and unchanged input express the stated obligations. Expecting a particular equal-key order can impose unstated stability. Oracle representation and applicability conditions therefore affect both fault detection and acceptance of legal behavior.

\subsection{Program Analysis and Hybrid Search}
Program analysis provides explicit structure that the model would otherwise need to infer from text. Dependency information can identify construction requirements; call relationships can constrain operation sequences; coverage can identify unexercised branches. CITYWALK and SAINT illustrate different uses of program structure for executable testing~\cite{U08,A02}. The source of structure matters: a graph recovered from implementation describes reachable operations, while a behavioral contract specifies what those operations should mean.

CODAMOSA combines search-based test generation with LLM-generated examples when coverage progress stalls~\cite{U04}. This division of labor is different from asking an LLM to generate an entire suite repeatedly. Search performs repeated local exploration, and the model supplies candidates that may open new regions. Fuzzing-driver and mutator generation extend the same principle to reusable generators~\cite{F05,F06,F07}. G2FUZZ reuses synthesized generators, while Scheduzz allocates execution resources across campaigns~\cite{F08,F10}. Model-call cost, artifact lifetime, and conventional search effort jointly determine the contribution of generation.

Hybrid comparisons need matched search engines, instrumentation, stopping rules, and execution budgets. Component ablations should isolate the model's contribution while accounting for conventional computation.

\subsection{Execution Feedback and Iterative Repair}
Feedback can localize malformed syntax, missing dependencies, or unexercised branches. An assertion failure is ambiguous: the test, implementation, environment, or requirement interpretation may be wrong. The acceptance decision must preserve this distinction.

CoverUp uses coverage and execution information in iterative Python test generation~\cite{U05}. TestGen-LLM filters proposed improvements through operational criteria that support regression-test integration~\cite{U07}. MuTAP uses mutation testing to guide generation~\cite{U06}. These systems demonstrate that feedback can be used for different decisions: repair a candidate, reject it, request another test, or redirect search. Recording only that a system is ``execution-guided'' loses this distinction.

Repair is an intervention on a particular part of the test. ChatTester and YATE refine near-miss tests, and YATE also discusses revising assertions against observed program behavior~\cite{U18,E09}. DCAware studies missing-oracle completion for executable regression prefixes and reports a self-repair trap in which iterative repair can favor easier-to-pass assertions~\cite{O13}. These accounts address different repair operations and settings. Their comparison should retain whether a repair changes syntax, setup, inputs, or the behavioral judgment, and whether the current implementation is assumed to define expected behavior.

For conformance, replacing an expected value with candidate output can erase the violation being tested. Setup repair and oracle revision therefore need different evidence. RESTestBench makes SUT-feedback quality an explicit evaluation condition~\cite{B05}. Paired repair branches from the same initial test can measure recovery and degradation, then validate the final suites independently.

\subsection{Agents, Collaboration, and Adversarial Generation}
Agent architectures assign responsibilities such as planning, code generation, execution, and review to separate components. TestAgent follows a workflow inspired by human testing and incorporates static knowledge and tool use~\cite{U09}. CANDOR and Nexus emphasize collaboration around test oracles~\cite{O06,O03}. The analytical question is what separation changes: available information, decision authority, search diversity, or execution budget.

ConVerTest generates code and test candidates and screens them using self-consistency, verification prompts, and dual execution agreement~\cite{O10}. It avoids requiring a pre-existing reference implementation, while still relying on generated implementations and their executions. VALTEST uses semantic entropy and a learned classifier to assess synthesized test validity~\cite{O11}. Consensus and predicted validity are additional acceptance signals; their reliability depends on shared errors, training labels, and the evaluation distribution.

Shared models and context can correlate agent errors. A checker querying an independent source contributes different evidence even when it uses the same model. Matched-budget controls should compare multiple roles with one combined-context agent, sampling and selection, and deterministic checking.

AdverTest pairs test and mutant agents in an adversarial generation loop~\cite{U10}. Such inference-time competition changes the current artifacts. It differs from reinforcement learning that changes model parameters across tasks, discussed in Section~\ref{sec:learning}. Both can expose weaknesses in an existing test suite, but their generalization claims differ. A stronger suite for the current target is evidence about search; a stronger frozen model on new tasks is evidence about learning.

\begin{table}[t]
\caption{Representative mechanisms and the interpretation of their feedback. The last column describes the immediate evidence supplied by the signal, not the full set of claims in each paper.}
\label{tab:mechanisms}
\small
\begin{tabularx}{\linewidth}{@{}p{.20\linewidth}YY@{}}
\toprule
Representative work & Main operation or feedback & Immediate interpretation \\
\midrule
SymPrompt~\cite{U03} & Path-oriented prompt decomposition & Guidance toward a structural target \\
CODAMOSA~\cite{U04} & LLM candidates during search plateaus & New material for continued exploration \\
CoverUp~\cite{U05} & Coverage and execution feedback & Reachability and artifact execution \\
MuTAP~\cite{U06} & Surviving mutants guide prompts & Discrimination within a mutation model \\
TestGen-LLM~\cite{U07} & Pass, stability, and coverage filters & Suitability for regression augmentation \\
Nexus~\cite{O03} & Candidate execution and agent discussion & Executed behavior plus model interpretation \\
AdverTest~\cite{U10} & Test--mutant interaction at inference & Adaptation to the current adversary \\
Bulbasaur~\cite{F07} & Branch-guided mutator synthesis & Reusable input exploration capability \\
\bottomrule
\end{tabularx}
\end{table}

Table~\ref{tab:mechanisms} links operations to immediate evidence. Information access and acceptance decisions explain their differences more precisely than strategy names alone.

\section{Learning Test Generators}
\label{sec:learning}
\subsection{Supervised Learning and Training Data}
Training changes the generator's behavior beyond one target or interaction. Supervised learning commonly uses code--test pairs or descriptions paired with tests. AthenaTest provides a historical example of learning unit-test generation with focal context~\cite{U11}. Such data teach framework syntax and common assertion patterns, but the training label also encodes a particular developer's testing choices. Multiple valid tests can exist for the same function, and a recorded test can be incomplete or redundant. Exact reproduction of a reference test is consequently a limited measure of learned testing ability.

Training corpora and task-specific adaptation supply complementary evidence. CAT-LM aligns code and test files during pretraining; UniTSyn uses language-server information to extract focal-code/test relationships across languages~\cite{T13,B09}. FuzzAug expands training data through coverage-guided fuzzing~\cite{T15}. exLong specializes in exception tests, while TOGLL and Tratto learn oracle construction~\cite{U16,O07,O09}. These examples distinguish learning repository correspondence, expanding observed inputs, and specializing a behavioral judgment. UniTSyn is principally a training-data resource, rather than another end-task evaluation benchmark.

Data filters determine the supervision: passing examples teach executability; discriminating examples additionally depend on the chosen faulty-program pool. Held-out projects and independently constructed faults test whether training transfers beyond that pool.

\subsection{Reinforcement Learning and Reward Meaning}
PyTester combines supervised learning and reinforcement learning for text-to-test generation, using automatic signals that include executable reference information~\cite{T01}. RLSQM learns from automatic feedback associated with unit-test quality~\cite{T12}. These approaches illustrate why the reward must be described in terms of the property being optimized. Syntax, execution success, structural coverage, maintainability, and semantic validity are related engineering concerns, but a reward for one does not directly establish improvement in another.

CURE makes the supervision distinction particularly clear. Its main co-evolutionary training setting omits ground-truth code supervision but pairs tasks with ground-truth unit tests, which help label generated programs and construct coder/tester rewards~\cite{T14}. Thus, absence of a reference implementation does not establish absence of reference supervision. Candidate programs, reference tests, and reference code must each be recorded, including changes across ablations.

Training access and inference access describe different dependencies. PyTester can generate from a description while training rewards execute a reference~\cite{T01}. Reference passing also assumes that the reference satisfies the requirement and that the test ignores incidental behavior.

\begin{table}[t]
\caption{Selected information-access configurations. Entries describe the cited configuration or stage; they are not claims about every variant or the underlying model's pretraining data.}
\label{tab:access}
\small
\begin{tabularx}{\linewidth}{@{}>{\raggedright\arraybackslash}p{.19\linewidth}YYY@{}}
\toprule
Configuration & Generation context & Additional evidence source & Consequence for interpretation \\
\midrule
TestGen-LLM~\cite{U07} & Existing tests and program context & Execution, stability, and coverage checks & Preserves observed regression behavior \\
AutoUAT / Test Flow~\cite{R03} & User stories, then scenarios and HTML & Human assessment in the industrial study & Design and implementation stages have different access \\
Nexus~\cite{O03} & Task information and agent interaction & Candidate implementation execution & Observations depend on candidate behavior \\
PyTester~\cite{T01} & Description and signature at inference & Reference execution during training & Inference access does not describe supervision needs \\
UTRL~\cite{T02} & Task-conditioned test generation & Reference and sampled programs during training & Adversarial learning retains a reference dependency \\
CURE, main training~\cite{T14} & Task-conditioned code/test generation & Ground-truth tests label generated programs & No ground-truth code does not mean no reference supervision \\
ConVerTest~\cite{O10} & Task and generated code/test candidates & Candidate execution and consensus & No pre-existing reference; candidate evidence remains \\
Requirements-oracle study~\cite{O12} & Requirements and minimal method structure & Researchers derive requirements from repair differences & Generator access differs from construction access \\
\bottomrule
\end{tabularx}
\end{table}

Table~\ref{tab:access} records stage-specific evidence, rather than assigning one black-box label per paper. Rewards should improve detection while constraining rejection of permitted behavior. Reference consistency is a practical proxy when alternative legal implementations are unavailable; the reference assumption remains explicit.

\subsection{Adversarial Learning and Self-Play}
UTRL learns a test generator through interaction with a code generator and uses reference and sampled programs in its training procedure~\cite{T02}. ATGen develops adversarial reinforcement learning for test-case generation with a changing adversarial curriculum~\cite{T03}. Sol-Ver studies self-play for solving and verifying code tasks~\cite{T04}. TCS separates stages aimed at reference-consistent tests and adversarial testing~\cite{T07}. CURE jointly evolves coder and tester behavior with reference-test-derived supervision in its main setting~\cite{T14}. Together, these approaches move test generation from imitating recorded tests toward learning from the behavior of generated programs.

Adaptive difficulty is valuable if it transfers beyond the current opponent. Evaluation should freeze the tester and use held-out tasks, independent error pools, and other code-model families. Matched-budget fixed or refreshed pools isolate interaction from additional erroneous examples. Agreement alone can preserve a shared misunderstanding; requirements, verified constraints, references, or human judgments anchor the labels.

\subsection{Learning Inputs and Learning for Downstream Use}
SAGE and LogiCase use grammar-based representations to connect input specifications to generated test inputs~\cite{T08,T09}. Their contribution concerns admissible and diverse inputs. An evaluator still needs a mechanism for deciding whether the resulting program behavior is correct. This division can be valuable because input validity has a different structure from expected-output reasoning and can sometimes be checked mechanically.

UTGen targets tests useful for debugging, whereas CodeRM trains and scales tests to support code reward modeling~\cite{T05,T06}. These objectives suggest an additional evaluation layer: the same test generator should be assessed both as a testing component and inside its intended downstream workflow. Improved code selection can arise from a better ranking signal even if some tests remain incorrect. Conversely, a semantically correct but weak test suite may have little effect on selection among already plausible candidates. Reporting both layers explains how the generated tests contribute to the final outcome.

\section{Benchmarks and Quality Evidence}
\label{sec:evaluation}
\subsection{A Quality Profile Rather Than a Single Score}
We organize test quality as a profile of complementary properties. \emph{Input validity} asks whether the input satisfies the task's domain constraints. \emph{Executability} asks whether the test can be built and run in the intended environment. \emph{Oracle correctness} asks whether its judgment follows the intended specification. \emph{Defect detection} asks whether it rejects relevant invalid behavior. \emph{Non-rejection} asks whether it accepts permitted behavior. Coverage, stability, maintainability, and cost add further dimensions needed for deployment.

Denominators differ: compilation can cover all attempts, whereas coverage and mutation often use retained tests. Report attempted, executable, accepted, and retained artifacts to expose filtering. Per-test and per-suite results are complementary, since one invalid assertion can cause an otherwise useful suite to reject a correct implementation.

\begin{figure}[t]
\centering
\includegraphics[width=\linewidth]{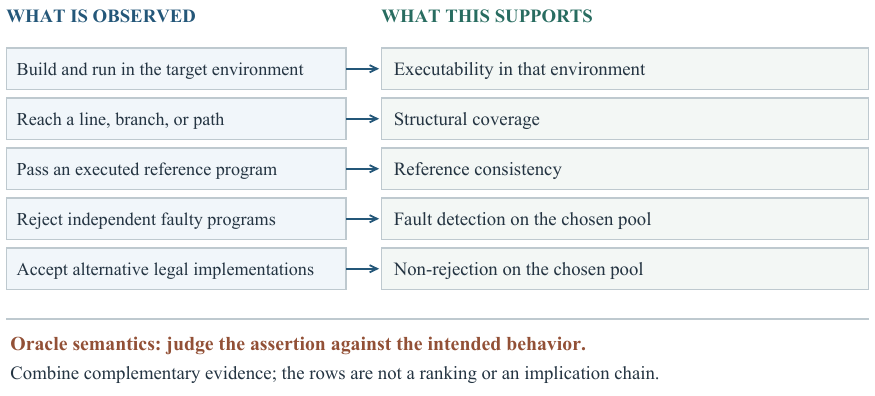}
\caption{Quality evidence supports different conclusions. Each arrow connects an observation to its immediate interpretation. Oracle correctness additionally depends on the intended specification and the validity of the evidence sources. These dimensions are complementary rather than successive levels of one score.}
\Description{Five separate arrows connect successful execution, structural reachability, reference passing, rejection of faulty programs, and acceptance of legal implementations to their corresponding bounded quality claims. No arrows connect these claims into a hierarchy.}
\label{fig:evidence}
\end{figure}

Figure~\ref{fig:evidence} connects each observation to its bounded interpretation. Coverage measures reach; mutation depends on its fault model and test validity; reference passing assumes a suitable reference. Assertion matching measures similarity, code-selection accuracy measures downstream utility, and repeated-run passing measures stability. These signals remain useful at their respective levels.

\subsection{Benchmark Families}
TestGenEval evaluates test generation and completion in real Python repositories~\cite{B01}. TESTEVAL emphasizes test generation for program coverage, including targeted structural objectives~\cite{B02}. Both are relevant to executable testing, but their task construction changes the context and the question being measured. Repository tasks expose dependencies and established tests; structural targets isolate whether generation can reach particular code regions.

SWT-Bench evaluates tests distinguishing buggy software from real-world fixes~\cite{B03}. Patch discrimination requires issue-related failure and acceptance of the fix, but does not characterize every legal implementation.

ULT/PLT studies test generation from real-world functions and examines contamination-related evaluation conditions~\cite{B04}. RESTestBench studies REST API tests generated from natural-language requirements, including requirement precision and SUT feedback~\cite{B05}. These benchmarks expose issues that small function tasks can underrepresent: complex context, ambiguity, dependencies between operations, and the quality of feedback used to revise a test.

Recent empirical work adds complementary evaluation dimensions. A study of real Python bugs compares LLM and human unit tests through fault detection~\cite{E01}. Research under software evolution studies test generation across semantic changes~\cite{E02}. Database testing work investigates flakiness in industrial and open-source settings~\cite{E03}. These studies broaden the target from initial generation quality to the behavior of tests on historical defects, changed software, and repeated executions.

Newer resources isolate further comparison conditions. TC-Bench selects a compact pool of erroneous contest programs through a code--test outcome matrix~\cite{B06}. Its discrimination and rank properties concern that observed pool. MultiFileTest moves evaluation to multi-file projects and distinguishes error-fixing configurations~\cite{B07}. TOGBench covers several oracle forms and reinserts generated oracles into Java tests for compilation, execution, and mutation assessment~\cite{B08}. The latter's repository uses ``false positive'' for erroneous programs that still pass; we retain that operational definition rather than equating it with rejection of legal implementations.

\begin{table}[t]
\caption{Benchmark selection should follow the intended claim. Rows identify primary emphases rather than exhaustive benchmark capabilities.}
\label{tab:benchmarks}
\small
\begin{tabularx}{\linewidth}{@{}p{.22\linewidth}YY@{}}
\toprule
Benchmark family & Useful question & Comparison condition to preserve \\
\midrule
TestGenEval~\cite{B01} & Can tests be constructed in repositories? & Existing tests, dependencies, generation versus completion \\
TESTEVAL~\cite{B02} & Can tests reach structural targets? & Target visibility and coverage denominator \\
SWT-Bench~\cite{B03} & Can tests expose an issue and accept its fix? & Issue information and buggy/fixed execution protocol \\
ULT/PLT~\cite{B04} & How do real functions and leakage conditions affect evaluation? & Dataset construction and model exposure assumptions \\
RESTestBench~\cite{B05} & How do requirements and feedback affect API tests? & Requirement precision and feedback validity \\
TC-Bench~\cite{B06} & Can tests distinguish selected erroneous solutions? & Candidate-pool construction and matrix definition \\
MultiFileTest~\cite{B07} & Can tests use multi-file project context? & Language, repository context, and repair configuration \\
TOGBench~\cite{B08} & Can generated oracles replace diverse Java oracles? & Oracle type, reinsertion, and metric definitions \\
\bottomrule
\end{tabularx}
\end{table}

The benchmark portfolio should follow the claim: repository usability, requirements reasoning, patch discrimination, and independent discovery require different evidence. Scores from unrelated protocols cannot be combined into a method ranking.

\subsection{Implementation State and Metric Interpretation}
Empirical studies increasingly separate tests generated from correct and buggy code. Work on incorrect-code influence and buggy-code misguidance examines how implementation errors affect generated judgments~\cite{E04,E06}. Replacing code with a generated docstring can change that influence, but the docstring remains code-derived~\cite{E06}. The requirements-oracle study instead evaluates requirement agreement separately from agreement with the SUT; its ten Lang-bug cases and repair-derived requirement construction bound the setting~\cite{O12}.

The same distinction qualifies proxy metrics. A replicability study reports that coverage and mutation can inform model comparison in regression settings where code is reasonably assumed correct, while their reliability for detecting existing bugs is weaker~\cite{E05}. Test Wars also illustrates that mutation, coverage, and fault-detection rankings can differ across LLM, search-based, and symbolic approaches~\cite{E08}. These observations support task-specific metric interpretation, rather than discarding structural metrics. The Java-oracle study uses developer-oracle comparison as a correctness proxy and mutation as a detection proxy~\cite{E07}; neither exhausts specification conformance.

Fuzzing adds a related failure-attribution problem. A crash from a generated driver can reflect an invalid API use rather than a reachable product defect. Google's OSS-Fuzz-Gen report examines false-positive mitigation using API constraints and realistic entry-point context~\cite{F09}. Valid input or driver construction, reproduced failure, and confirmed defect should therefore be separate stages of the reporting funnel.

\subsection{Independent Oracle Assessment}
The requirement can be viewed as defining a set of permitted implementations. For an illustrative evaluation, let $\mathcal{P}^{+}$ be independently validated implementations that satisfy the requirement and let $\mathcal{P}^{-}$ be independently constructed faulty implementations. For a test suite $T$, define
\begin{equation}
\mathrm{RejectLegal}(T)=\frac{\sum_{p\in\mathcal{P}^{+}}\mathbf{1}[T(p)=\mathrm{fail}]}{|\mathcal{P}^{+}|},\qquad
\mathrm{DetectFault}(T)=\frac{\sum_{p\in\mathcal{P}^{-}}\mathbf{1}[T(p)=\mathrm{fail}]}{|\mathcal{P}^{-}|}.
\end{equation}
The two quantities measure complementary behavior on finite evaluation pools. They require a nonempty pool of each kind, valid execution environments, and trustworthy pool labels. They are an evaluation recommendation rather than measurements extracted from the reviewed studies. Reporting them together exposes suites that detect many faults by also rejecting permitted behavior. Neither quantity alone proves specification-wide correctness.

Alternative correct algorithms and permissible output choices test whether an oracle encodes unstated implementation details. Faults withheld from training, prompting, and feedback test independent detection. Where executable pools are unavailable, expert scenario assessment should record traceability, applicability, expected behavior, omissions, and assessor disagreement; textual similarity alone does not establish detection.

\subsection{Comparability, Leakage, and Cost}
Comparisons need matched information contracts: requirements-only generation, reference access, and candidate-feedback repair are different settings. Training data, retrieval, existing tests, and evaluation artifacts can cross intended boundaries. Contamination controls should identify the paths they address.

Cost should include model calls, input and output tokens, conventional executions, environment preparation, and human intervention. For trained generators, training cost and deployment cost answer different questions and should be reported separately. For hybrid fuzzers, execution throughput and the reuse of generated infrastructure matter. For agent systems, total calls across all roles matter. An equal-token comparison and an equal-wall-time comparison may favor different designs, so both can be informative when resources permit.

Generation randomness and execution nondeterminism are separate sources of variance. Project-level results, repeated generation, repeated execution, failed attempts, and stopping rules make uncertainty and selection effects assessable.

\section{Practical Integration and Research Agenda}
\label{sec:agenda}
\subsection{Choose a Workflow from the Testing Objective}
Workflow selection starts with the behavioral decision, available evidence, and missing artifact. Regression augmentation can use existing tests and coverage, as in TestGen-LLM~\cite{U07}. Conformance needs requirement-backed judgments; fuzzing needs valid generators and failure attribution; interactive code generation can use user-confirmed examples, as in TiCoder~\cite{T11}. Assertion provenance connects these choices to review and maintenance.

\subsection{Industrial Adoption and Human Revision}
Industrial evidence includes different acceptance decisions. TestGen-LLM studies improvement of existing tests through operational filtering~\cite{U07}; the acceptance-testing case study evaluates both generated scenarios and their conversion into executable scripts~\cite{R03}. RAGTAG and system-level test-specification generation additionally involve domain interpretation and assessment~\cite{R01,R02}. ASTER adds dependency handling and mocking, while AToMIC connects industrial tickets to mobile acceptance scripts~\cite{U12,R06}. The OSS-Fuzz-Gen report adds failure-validation work to the adoption burden~\cite{F09}. A scenario judged useful by an engineer and a script accepted into a regression workflow are distinct outcomes.

A deployment evaluation should retain these stages: generated, reviewed, revised, executed, and adopted. Human effort includes interpreting requirements, supplying missing context, correcting assertions, and maintaining the accepted test. Reporting the time and rejection reasons at each stage would connect the artifact's technical quality to its actual workflow value. The reviewed examples motivate these dimensions; this synthesis does not pool their heterogeneous adoption measures.

\subsection{Oracle Provenance and Independent Validation}
Linking an assertion to a requirement clause, validated example, formal property, or regression observation could guide acceptance and repair. The research question is whether checked provenance improves oracle quality and maintenance. A matched-budget comparison should measure unsupported assertions, rejection of legal implementations, independent detection, and review time. Plausible explanations without verified evidence links do not establish that benefit.

\subsection{Feedback That Preserves Valid Tests}
Repair should recover invalid tests while preserving valid judgments. Branching from the same initial artifact allows separate measurement of recovery and regression, followed by independent suite evaluation. Controls can compare setup-only repair, unrestricted repair, no semantic repair, and trusted-reference checking. The information and budget must match; improved passing alone cannot determine whether assertions were corrected or weakened.

\subsection{Generalization in Learned and Adversarial Testers}
Adversarial curricula offer difficult errors~\cite{T02,T03,T07}, but transferable capability requires a frozen tester evaluated on new opponents, repositories, and specification styles. Suite rewards also obscure which tests help or mislead. Comparing aggregate rewards with verified test-level feedback can test the value of finer credit assignment, while retaining validity constraints and independent detection.

\subsection{Repository Scale and Lifecycle Evaluation}
TestGenEval, CITYWALK, and service-level testing provide repository-oriented starting points~\cite{B01,U08,A02}. Lifecycle evaluation should additionally label whether a change preserves semantics or deliberately changes requirements. Evolution and database-flakiness studies motivate retaining prior tests and measuring execution stability, maintenance effort, and unintended-regression detection~\cite{E02,E03}. Version-aware retrieval should be compared with unrestricted context reuse, since existing tests can preserve brittle setup or obsolete obligations.

\subsection{Reproducible Evidence and Efficient Allocation}
Reproducibility packages should expose the path from request to retained suite. Relevant records include model and prompt versions, context selection, generated candidates, tool feedback, repairs, rejection reasons, and final evaluation inputs. This enables analysis of why one method succeeds and whether its filters hide a large failure burden. For training methods, data construction and reward implementation are equally important.

Budget allocation is another cross-cutting research question. A fixed budget can be spent on generating more candidates, checking oracles, executing additional faults, or resolving uncertain requirements. Dynamic scaling in CodeRM~\cite{T06} and reusable fuzzing infrastructure~\cite{F05,F06,F07} illustrate different responses to this allocation problem. A broader study could compare these investments under a common deployment objective, such as maximizing independent detection while bounding invalid-suite rate and latency.

\section{Threats to Validity}
The collection supports a qualitative map of research directions rather than estimates of their prevalence. Web-based discovery, English-language emphasis, and the stated source restrictions can omit relevant studies. Performance, accessibility, concurrency, and distributed-system testing received less focused coverage than functional testing and fuzzing. Selected neural predecessors provide context but do not constitute a complete history of learned test generation.

Inspection depth varies: abstract-level evidence supports broad mechanism descriptions; context permissions, reward definitions, and denominators require fuller coding. We do not pool numerical gains. Multi-database retrieval, complete variant extraction, and independent coding remain necessary for a systematic extension. The framework has not undergone inter-rater validation, and multi-stage methods can occupy several categories. Preprint revisions, changing model services, and publication bias further limit the dated synthesis.

\section{Conclusion}
LLM-based test generation spans scenarios, executable suites, oracles, and reusable fuzzing infrastructure. We connected these artifacts to their information sources, generation mechanisms, and evaluation assumptions. The synthesis separates executable construction from oracle justification: coverage, reference consistency, patch discrimination, independent detection, and acceptance of legal behavior supply complementary evidence. Making information access, evidence provenance, and cost explicit supports more interpretable comparisons and guides research on generators that remain useful across repositories, adversaries, and software changes.

\section*{Data Availability}
The accompanying materials are available at \url{https://github.com/dududuck00/llm-test-generation-survey}. They comprise the 95-record source register, search and inspection notes, supplemental provenance, synthesis outline, bibliography, manuscript sources, and editable conceptual figures. The register preserves primary-source links and inspection-depth annotations. No experimental dataset or reproduced performance results are claimed.

\clearpage
\bibliographystyle{ACM-Reference-Format}
\bibliography{references}
\end{document}